
\input phyzzx
\def\ls#1{_{\lower1.5pt\hbox{$\scriptstyle #1$}}}

\input phyzzx

\def\SCIPP{\centerline {\it Santa Cruz Institute for Particle Physics}
  \centerline{\it University of California, Santa Cruz, CA 95064}}
\overfullrule 0pt
\titlepage
\Pubnum{SCIPP 93/17}
\date{June, 1992}
\vskip3cm
\title{{
Some Questions of Flavor in Supersymmetry}
\foot{Invited Talk Presented at Conference SUSY93, Northeastern
University, March 1993}}
\foot{Work supported in part by the U.S. Department of Energy.}
\author{Michael Dine
}
\address{}
\SCIPP
\vskip.5cm
\vbox{
\centerline{\bf Abstract}

\parskip 0pt
\parindent 25pt
\overfullrule=0pt
\baselineskip=18pt
\tolerance 3500
\endpage
\pagenumber=1
\singlespace
\bigskip


We consider certain naturalness questions in supersymmetric theories.
Various suggestions which give rise to squark degeneracies are
reviewed.  A stringy scenario, discussed by Kaplunovsky and Louis,
is the only one which leads to complete degeneracy of squarks and
sleptons at the high scale.  Alternatives include the possible
existence of a gauged non-Abelian
horizontal symmetry, broken at some scale,
and theories in which the ``messengers" of supersymmetry breaking
are gauge interactions.  A model of the latter type is described,
in which supersymmetry is {\it dynamically} broken at TeV energies.
Models of this type can solve many of the naturalness
problems of supersymmetric theories, and predict a rich phenomenology
at SSC energies.
}
\singlespace
\bigskip
\chapter{Introduction}

The hierarchy problem suggests that there is new physics at TeV
energy scales.  The candidates which we know for this
physics are  supersymmetry and
technicolor, though we should keep in mind that there may be something
else (we have no good ideas for understanding the vanishing
of the cosmological constant, for example).   These are very attractive
ideas, but both face potentially serious difficulties associated with
problems of flavor.

\REF\randall{L. Randall, MIT-CTP-2112 (1992); H. Georgi, HUTP-92-1037
(1992).}
It is usually said that technicolor theories fail when they confront
the issue of flavor changing neutral currents.  (For recent efforts
to solve this problem, see ref.~\randall.)  It is also often said
that supersymmetry does not suffer from such difficulties.
In particular, if squarks of different flavors are approximately degenerate,
then the contribution of diagrams containing new supersymmetric particles
are small.  The degree of degeneracy required is quite severe,
however.  If the scale of supersymmetry breaking is, say, 300~GeV,
then from the real part of $K$--$\bar K$ mixing one has
$${\delta m^2 \over m_{\rm SUSY}^2} < 10^{-2}\ .\eqn\degeneracyconstraint$$
As one increases the typical SUSY-violating mass, $m\ls{\rm SUSY}$, this limit
only improves linearly.

\REF\il{L. Ibanez and D. Lust, Nucl. Phys. {\bf B382}
(1992) 305.}
It is often argued that such a degeneracy is natural, at least at the Planck
mass, $M_p$.  In most considerations of the minimal supersymmetric
standard model (MSSM), for example, one assumes that at the high scale,
the soft breakings have the structure:
$$V_{soft}=m_{\rm SUSY}^2\sum \vert \phi_i \vert^2 + A \lambda_{ijk} + \dots
\eqn\genericsoft$$
(where $\lambda$ denote the cubic couplings in the superpotential).
We will refer to the assumption that the $\phi\phi^*$ type mass
terms are equal as the assumption of ``degeneracy;"  the assumption
that the cubic soft-breaking terms are proportional to the corresponding
terms in the superpotential we will refer to as the assumption of
``proportionality."
In the framework of supergravity theories, one frequently
hears the assertion
that such a universality of couplings is reasonable; gravity, after all,
is
universal.  But this argument, as it stands, is specious.
In general relativity, matter couples universally
to gravity as a consequence of a symmetry principle (and
the fact that only the lowest dimension operators are
important).  In a supergravity theory, on the other
hand,
no symmetry that we observe at low energies requires degeneracy
or
proportionality.  In string theory, for example, there is no reason to expect
such relations to hold generically, and Ibanez and Lust have
explicitly verified
that they do not.\refmark{\il}
There are other flavor problems in supersymmetry as well:
the neutron and electron electric dipole moments ($d_n$ and $d_e$)
severely constrain $CP$-violating phases in the soft-breaking terms.

In this talk, we will see that there
are in fact a number of ways (at least three) in which an adequate degree
of degeneracy and proportionality can arise at tree level in supersymmetric
theories.
Even before describing these,
it should be stressed that the situation is not nearly so
problematic as in the
case of technicolor.  In particular, if one assumes degeneracy and
proportionality
at the high scale, then ignoring the small Yukawa couplings
(i.e., all but those of the $t$ and perhaps the $b$
quarks)
there {\it is} a large, approximate flavor symmetry. This symmetry insures that
radiative
corrections to these relations are very tiny (assuming a cutoff
of order $M_p$),
and that flavor-changing processes are readily suppressed.

\REF\cargese{G. 't Hooft, in {\it Recent Developments
in Gauge Theories}, G. 't Hooft et al., Eds., (Plenum,
New York, 1980).}
To understand the problem in more detail, consider a general $N=1$
supergravity model, with supersymmetry broken in a hidden sector.
Denote the hidden sector fields by $z$, and the visible sector fields
by $y$.  The general supergravity lagrangian (up to terms with
two derivatives) is specified by three functions:  the Kahler potential,
$K(\phi,\phi^*)$, the superpotential, $W(\phi)$, and a function
$f(\phi)$ which describes the gauge couplings. In general, we can
write, after suitable rescalings of the fields,
$$K= k(z,z^*)+\sum_i y_i^*y_i + \ell_{ij}(z,z^*)y_i y_j^*+\dots
\eqn\generalkahler$$
If one writes the general potential in terms of $K$ and $W$, one
sees that the condition for degeneracy and proportionality is that
$\ell_{ij} \propto \delta_{ij}$.
As we have noted, no symmetry enforces this, and there is no reason, in
general,
to expect degeneracy.
This condition is certainly not satisfied in any
generic sense in string theory.\refmark{\il}
If we think in terms of 't Hooft's naturalness
criterion,\refmark{\cargese} it is clear that we do
not expect degeneracy between squarks and
sleptons to hold to better than ${\cal O}(\alpha/\pi)$,
and between squarks of different flavors than to
better than order Yukawa couplings.

\REF\vadim{V.S. Kaplunovsky and J. Louis,
CERN-TH. 6809/93 UTTG-05-93.}
\REF\louisetal{R. Barbieri, J. Louis and M. Moretti, CERN-TH.6856/93.}
\REF\dks{M. Dine, A. Kagan and S. Samuel,
Physics Letters {\bf B243} (1990) 250.}
\REF\dkl{M. Dine, A. Kagan and R. Leigh,
 SCIPP-93-05.}
\REF\nirseibergb{Y. Nir and N. Seiberg, RU-93-16.}
\REF\dinenelson{M. Dine and A. Nelson,
SCIPP-93-03.}
I know, however, of  four situations which can give rise to a significant
degree
of degeneracy and proportionality:
\noindent
\pointbegin
Kaplunovsky and Louis\refmark{\vadim}
have  recently pointed out that given certain assumptions about
SUSY breaking,
string theory {\it can} give rise to a significant level
of degeneracy.  In particular, if the auxiliary field associated with
the dilaton supermultiplet, $F_S$, is much larger than that associated with
other moduli, $F_M$, then all squarks and sleptons are indeed degenerate
at tree level.  These authors argue that it is hard to understand how this
could
come about, but I would suggest that
given how little we understand about
SUSY breaking in string theory\foot{Indeed, one can well
argue that there may be no sensible breaking of
susy in string theory at weak coupling; certainly, none of
the scenarios which have been proposed satisfy all of the
conditions enumerated in ref.~\vadim.}
this may be a significant clue to
string dynamics.  The phenomenology of this scenario has been
considered in ref.~\louisetal.
\noindent
\point
If at the high scale the scalars are very light compared to the
gauginos, some degree of degeneracy results.\refmark{\dks}
\noindent
\point
Perhaps there really is a flavor symmetry at the high scale.
It must be broken in such a way that there are large non-degeneracies
among fermions, yet a high degree of degeneracy among scalars.
In the next section, I will focus on the possibility that these horizontal
symmetries
are non-abelian.\refmark{\dkl}
Since I gave this talk, the possibility of achieving
this result with Abelian symmetries has been
explored.\refmark{\nirseibergb}
\noindent
\point
Perhaps supersymmetry is broken at a rather low scale, and supersymmetry
breaking is fed to ordinary fields through gauge interactions.
In that case; squarks with a given set of gauge quantum numbers will
be approximately degenerate.  Recently, models of this type where
supersymmetry breaking arises {\it dynamically} have been
constructed.\refmark{\dinenelson}  These models will be
reviewed briefly in section 3.

\vskip3pt
Only the first of these approaches leads to a MSSM
phenomenology of the type
which has been so widely explored recently.  In the flavor
symmetry scheme, one does not expect much degeneracy between
squarks or sleptons with different gauge quantum numbers, nor
between particles in the first two and the third generations.
In the case that supersymmetry breaking is fed by gauge interactions,
one expects masses to be roughly proportional to appropriate gauge
couplings.  All of this suggests that we should be open to a broad
range of possibilities.  The rest of this talk will focus on the
third and fourth approaches to solving the degeneracy problem.

\chapter{Non-Abelian Flavor Symmetries and Squark Degeneracy}

An obvious solution to the problem of squark degeneracy is to suppose
that there is an underlying, gauged non-Abelian flavor symmetry.
Equally obvious is that this symmetry must be badly broken in order
to account for the fermion mass matrix,
$m_F$.  The issue is how much degeneracy
is possible with a realistic $m_F$.
We will adopt a rather simple-minded approach, and will not attempt
to {\it explain} $m_F$.  We will follow two basic rules:
\item{1.}  The horizontal symmetry will be a gauge symmetry.
(More definitely, we will require that any continuous
horizontal symmetry be a gauge symmetry.)  This is consistent with the
dogma that the only exact continuous symmetries should be gauge symmetries.
\item{2.}  We will impose 't Hooft's naturalness criterion.\refmark{\cargese}
We will insist that couplings (or sets of couplings) are only small if the
theory
becomes more symmetric in the limit that those couplings go to zero.  This
works two ways:  new supersymmetric couplings can only be small if the
theory becomes more symmetric in that limit; some of them {\it must} be
small if one is to understand the smallness of corresponding Yukawa couplings.

\vskip3pt
\REF\strominger{A. Strominger and E. Witten, Comm. Math. Phys.
{\bf 101} (1985) 341.}
\REF\fayet{M. Dine, N. Seiberg, and E. Witten, Nucl. Phys. {\bf B289}
(1987) 589.}
The first question we must address is the scale of flavor symmetry breaking.
Here I will
assume that the breaking scale is near (in fact slightly below) $M_p$.
Obviously it is of interest to explore breaking at much lower scales.
The high energy scenario can be motivated in a stringy way.  It is
well-known that as one explores the moduli space of string
compactifications, one frequently finds points of enhanced symmetry
(e.g., the SU(3) symmetry of the simple $Z_3$ orbifold).  Suppose
that the horizontal symmetry is of this type.  Then at this point
there are moduli, i.e., fields with
no potential,
which transform under the symmetry.  We will denote these
fields generically by ${\Phi}$.
The vev's of these fields break flavor.  They might also break $CP$
spontaneously.\refmark{\strominger}
Their natural values are $0$
(corresponding to unbroken symmetry) or $M_p$.  In what
follows we will suppose
$${\VEV{\Phi} \over M_p} \sim {1 \over  10}-{ 1 \over 100}\ .
\eqn\modulivevs$$
Such values are certainly plausible.  For example, in string theory,
where Fayet-Iliopoulos terms are frequently generated at one
loop, the $\Phi$ vev's might be expected to be of order
$\alpha$.\refmark{\fayet}

As an example, suppose that one has an SU(2) flavor symmetry,
and that the light quarks and leptons lie in doublets.
In particular, suppose that there are left-handed and right-handed
doublets
$Q^a ,\bar u^a , \bar d^a$
and singlets
$Q_s , \bar u_s, \bar d_s$.
The ordinary Higgs particles are assumed to be singlets of SU(2)$_H$,
while the moduli are assumed to be a set of $N$ doublets,
$\Phi_i^a,i=1\dots N$.
Consider first how the fermion mass matrix arises in this framework.
The superpotential just below $M_p$ contains
dimension-four terms:
$$W_q= \lambda_1 \epsilon _{ab} Q_a \bar d_b H_1 + \lambda_2 \epsilon _{ab} Q_a
\bar u_b H_2 + \lambda_3 Q_s \bar d_s H_1
+ \lambda_4 Q_s \bar u_s H_2\ . \eqn\wdimensionfour$$
These give rise to SU(2)$_H$-symmetric terms in the mass matrix.
Clearly we need to assume that $\lambda_1$ and $\lambda_2$
are small in order that the $u$ and $d$ quarks be sufficiently light
(this might be arranged by means of a discrete symmetry).
SU(2)$_H$-violating terms arise at the level of dimension five
and dimension six operators:
$$
{1 \over M_p}(\lambda_5^i \epsilon _{ab}\Phi_a^i  Q_b \bar d_s H_1
+ \lambda_6^i \epsilon _{ab}\Phi_a^i  Q_s \bar d_b H_1)
+{1 \over M_p^2}(\lambda_7^{ij} \epsilon _{ab}\epsilon _{cd}\Phi_a^i \Phi_c^j
Q_b \bar d_d H_1 +.....)\ .
\eqn\dimfiveandsix$$
Note that the charmed-quark mass must arise from these
operators, and is thus of order $(\Phi/M_p)^2$,
so $\Phi/M_p$ can't be much smaller than $0.1$.

Now consider soft-breaking terms.
The breaking of the squark degeneracy can also be
understood in terms of the effective action at scales
slightly below $M_p$.  This lagrangian contains
terms dimension-four, soft-breaking terms which
give SU(2)$_H$-symmetric contributions to the squark
mass matrices:
$$V_{soft}= m_1^2 \vert Q_a \vert^2 + m_2^2 \vert Q_s \vert^2+
m_3^2 \vert \bar u_a \vert^2 + m_4^2 \vert \bar u_s \vert^2 + ...$$
$$+A_1 \lambda_1 Q \bar d  H_1+A_2 \lambda_2 Q \bar u H_1 + .... + h.c.,
\eqn\softterms$$
Here, $m_\Phi$ and $A_i$ are of order $m\ls{\rm SUSY}$.
Breaking of the symmetry
will arise through terms of the type
$$\delta V^2_{soft}= {m_{\rm SUSY}^2 \over M_p}(\gamma _1
\Phi _1 Q Q_s^* + ...)
+ {m_{\rm SUSY}^2 \over M_p^2}(\gamma^{\prime}_1
\Phi_1 Q \Phi_2 Q^* + ...) \eqn\nonrensoftbil$$
and
$$\delta V^3_{soft} =  {m\ls{\rm SUSY}\over M_p}
\lambda_5^1 Q \bar d_s  H_1 (\eta_1 \Phi_1
+  \eta _2 \Phi_2+ \eta_3 \Phi_2^*)$$
$$ +{m_{\rm SUSY}\over M_p^2} \lambda_7^{11} Q \bar d H_1
(\eta '_1 \Phi_1 \Phi_1 +\eta '_2
 \Phi_1 \Phi_2 + \eta_3^{\prime} \Phi_1 \Phi_2^*)+...
 \eqn\nonrenormsofttril$$
We have omitted SU(2)$_H$ indices on $Q$, $\bar u$, $\bar d$, but terms
with all possible contractions should be understood.  Here $\gamma$,
$\gamma^{\prime}$, $\eta$ and $\eta^{\prime}$ are dimensionless
numbers.

By 't Hooft's naturalness criterion,\refmark{\cargese}
many of these
couplings should not be much less than one; the theory does not become
any more symmetric if these quantities vanish.
As a result, the generic symmetry-violating terms in the first two
generations are of order $(\Phi/M_p)^2 \sim 10^{-2}$.
This is by itself just barely enough to adequately suppress
the real part of $K$--$\bar K$ mixing.  Many of the couplings here,
however, can (and should!) be small by 't Hooft's criterion,
particularly off-diagonal couplings.  (One might imagine suppressing
these by imposing further discrete symmetries.)
As a result,
there is no difficulty with flavor-changing neutral currents.

If the phase of $\Phi^a$
is the origin of $CP$-violation, many of the new
supersymmetric contributions to $d_n$ and $d_e$ are automatically
suppressed.  Complex terms in the gluino mass
matrix must arise from terms in the function $f$ involving
$\Phi^a$; by gauge invariance, these terms are at least
quadratic in $\Phi$, and thus suppressed by two orders of
magnitude.  Similar remarks apply to the $A$ parameter.
However, in the present framework, complex, off-diagonal
terms can also arise in the squark mass matrices, and one
must make sure that these are adequately suppressed.  This
poses no more difficulty than for the $K$--$\bar K$ system.

\REF\nirseiberga{M. Leurer, Y. Nir and N. Seiberg, Rutgers preprint RU-92-59,
(1992).}
Clearly we have only scratched the surface of this subject.
Most importantly, one would like to consider this problem
in a framework which addresses the origin of quark and lepton masses,
for example as in the work of ref.~\nirseiberga.
In any case, certain predictions seem likely to emerge
from any framework using non-abelian horizontal
symmetries to solve the squark degeneracy problem.
Because the top quark mass is so much larger than the
others, one expects to encounter some sort of SU(2)-type
structure.  So one expects that
left and right-handed squarks will lie in
nearly degenerate doublets; there will be no degeneracy
between the first two generations and the third,
or between left and right.

\chapter{Dynamical Supersymmetry Breaking at Low Energies}

Dynamical supersymmetry breaking (DSB), needless to say,
has the potential to address a set of naturalness issues
which go well beyond the squark degeneracy problem.
It has the potential to {\it explain} the hierarchy
in terms of small, $e^{-c/g^2}$ effects.
In addition, soft breakings and other
effects should be calculable.  In this section, I will briefly describe
a model with
\noindent
\pointbegin
DSB (at a scale of order $100$'s of TeV).
\noindent
\point
SUSY breaking fed down to ordinary particles
by gauge interactions (leading to sufficient degeneracy
and proportionality for FCNC's).
\noindent
\point
Natural SU(2) $\times$ U(1) breaking; the low
energy particle content, however, is necessarily different
from that of the MSSM.

\noindent
While these features are certainly wonderful, the model
does have certain drawbacks:
\noindent
\pointbegin
It is complicated, and won't win any beauty contests.
\noindent
\point
The model possesses one potentially dangerous ``axion."
Provided a certain condition on couplings is satisfied, the
mass of this axion can be $\sim~100$~MeV.

\noindent
Neither of these problems is in any obvious sense generic.
Hopefully, they reflect the fact that we have not yet been
clever enough, and someone will soon construct more attractive
models with all of the good features listed above.
In any case, these theories certainly provide an
``existence proof," and a framework in which to study
the phenomenology of such low energy breaking.
I believe that such a scheme provides an interesting
alternative to that of the usual MSSM.

\REF\wittendsb{E. Witten, Nucl. Phys. {\bf B188} (1981) 513;
Nucl. Phys. {\bf B202} (1982) 253.}
\REF\dsbqcd{I. Affleck, M. Dine, and N. Seiberg, Nucl. Phys. {\bf B241}
(1984) 493.}
I do not have space here to fully review the issues
involved in DSB, or to describe the model in detail,
so instead I would like to just focus on some important
features.  The problem of DSB was first clearly posed
by Witten.\refmark{\wittendsb}
He pointed out that because of the non-renormalization
theorems, supersymmetry breaking is necessarily non-perturbative
and small at weak coupling.  He elucidated several conditions
for DSB to occur.  First,
one requires
a massless fermion to play
the role of the Goldstone fermion.  In many instances,
he argued, the issue is to show that a superpotential
is generated for the light fields.  Second, the ``Witten
index" must vanish;
For several interesting theories,
he could compute the index and showed that it was non-zero.

Subsequently,
it was shown that superpotentials
{\it are} generated in many               \unskip\break
theories.\refmark{\dsbqcd}
An example is provided by an SU(2) gauge theory with a single
quark flavor (corresponding to two chiral doublets, $Q$ and564$\bar Q$).
Classically, such a theory has a set of degenerate
vacua, with
$$Q=\left ( \matrix{v \cr 0} \right ) = \bar Q\ . \eqn\sutwoflat$$
For large values of $v$, the gauge symmetry is completely
broken and the theory is weakly coupled.  There is
one light field in the vacuum; it can be written as
$\Phi = \bar Q Q$, where it is understood
that the fields are to be expanded in small fluctuations about
their vacuum expectation values.  A straightforward
instanton calculation shows that a superpotential is generated
for $\Phi$,
$$W_{np}= {\Lambda^5 \over \Phi}$$
where $\Lambda$ is the usual renormalization group invariant
scale parameter of the theory.

While this example illustrates the non-perturbative
breakdown of the non-renormalization theorems, it does not
lead to a phenomenologically acceptable model, since the potential
tends to zero for large $v$.  If one adds a mass term, one finds
that there are two supersymmetric ground states (consistent
with the index).

These features turn out to be generic.  In order to obtain
DSB with a ``nice" vacuum, one finds that one must
satisfy two conditions (the ``Seiberg criteria"):
\item{1.}
The classical theory must have no flat directions
\item{2.}  The theory must possess a spontaneously broken
global symmetry.

\vskip3pt\noindent
It is easy to understand these criteria:  if SUSY is unbroken,
the Goldstone boson has a scalar partner, which
parameterizes a set of flat directions; by assumption
these don't exist.  This criterion is
admittedly heuristic, but it works in all known examples.

\REF\macintire{M. Dine and D. MacIntire,   Phys. Rev. {\bf
D46} (1992) 2594.}
\REF\weakcoupling{M. Dine and N. Seiberg,
Phys. Lett. {\bf 162B} (1985) 299;
and in {\it
Unified String Theories}, M. Green and D. Gross, Eds. (World Scientific,
1986).}
\REF\dsbphen{I. Affleck, M. Dine and N. Seiberg, Nucl. Phys. {\bf B256}
(1985) 557.}
If one wants to do phenomenology with such
models, there are two approaches one might consider.
\item{1.}
One might use such theories as hidden sectors for
supergravity (or superstring) theories.   The main
problems with such schemes lie in obtaining gaugino
masses,\refmark{\macintire}
and in trying to solve the other naturalness
problems of supersymmetric theories.  (Of course,
one might try to solve these using flavor symmetries
such as described in the previous section.)
In the case of superstrings, one encounters the usual
dilaton problem:  any potential generated
for the dilaton will tend to zero at weak
coupling.\refmark{\weakcoupling}

\item{2.}  Alternatively, one can consider low energy
breaking.\refmark{\dsbphen}
In this case, one tries to gauge a global symmetry of the
model, and identify it with the standard model gauge group.
Then gauge loops will give rise to squark, slepton and
gaugino masses.  This has the desirable feature that to
a very good approximation, squark and slepton masses
depend only on gauge quantum numbers, and there is
great suppression of flavor-changing processes.  However,
model building along these lines runs into a variety
of serious problems.  First, the simplest models exhibiting
DSB with a large enough flavor symmetry possess very
large gauge groups, and as a result, QCD is violently
non-asymptotically free.  Also, because of the presence of
spontaneously broken global symmetries, there are
typically unacceptable  axions and
Goldstone bosons.

\vskip3pt
\FIG\flowchart{Strategy for building a model with DSB
at low energies}
Here I would like to describe a solution to these problems.
The idea is simply to interpose an extra set of interactions
between the sector of the theory which breaks DSB
and the ordinary fields.  In other words,
there is a set of fields which are responsible
for breaking DSB; we will
refer to the associated gauge interactions as ``supercolor."
  Some of these fields carry an additional
(gauged) quantum number called ``R-color."  There are
a second set of fields, which I will refer to as ``straddlers,"
which carry both R-color and ordinary SU(3) $\times$ SU(2)%
$\times$ U(1) quantum numbers.  Finally, there are the
usual matter fields.  The ``straddlers" gain mass
as a consequence of $R$-color interactions.  Superpartners
of ordinary fields gain mass by emitting SU(3) $\times$ SU(2)%
$\times$U(1) gauge fields, which couple to the straddlers.

This approach is able to solve both of the problems listed
above.  First, because the $R$-color group need not be
so large, the contribution of the straddlers to the QCD
beta function need not spoil asymptotic freedom (in fact,
in the model of ref.~\dinenelson, it is possible to achieve
the usual unification of couplings).  Second, provided
the $R$-color interactions are sufficiently strong,
they can give a sufficiently large mass to the axion which
arises from the DSB sector.

There is not space here to review the model of ref.~%
\dinenelson\ in detail.  In that paper, a model
is analyzed with

\pointbegin
A sector whose full non-perturbative potential
possesses a (local) minimum with DSB.

\point
At this minimum, there is an unbroken SU(3)
(which plays the role of $R$-color) and
a broken SU(3), under which the straddlers transform.

\point
At one loop, the straddlers obtain supersymmetry-breaking
masses.  These masses are negative, but analysis of the
potential shows that SU(3) $\times$ SU(2) $\times$ U(1)
is unbroken for a range of parameters.  $R$-color is unbroken
as well; because all of the straddlers gain mass
at this stage, $R$-color can quickly become strong, giving
mass to a dangerous axion which arose at the first stage
of symmetry breaking.

\point
Below the scale of the straddler masses, it is necessary to
integrate out these fields.  In principle, one must compute
three loop graphs to obtain the masses of ordinary squarks
and sleptons.  However, these masses are proportional
to a log of the supercolor scale over the straddler mass.
This logarithmic term is easily isolated.  One finds that
its coefficient is positive, and that squark and lepton
masses are given by an expression of the form:
$$\tilde m_2 = \sum_i C_F^{(i)}{\alpha_i \over
\pi}^2 m_{\rm SUSY}^2\eqn\scalarmassformula$$
where $m_{\rm SUSY}^2$ is a common mass parameter,
and the sum is over the standard model gauge groups.

\point
Gaugino masses are proportional to $(\alpha_i /\pi) m\ls{\rm SUSY}$
(with a non-universal coefficient).

\point
SU(2) $\times$ U(1) breaking requires additional fields in
the theory.  The problem is that global discrete symmetries
of the model forbid an $H_1 H_2$ term in the potential.
Thus one must at least add a gauge singlet.  However,
if this is all one adds, only the Higgs field which couples
to the top quark can gain a negative mass-squared, and
the model necessarily contains an unacceptably light Higgs.
This problem can be solved by adding an additional set
of mirror fields, with sufficiently large couplings to the
singlet.  In this case, the singlet obtains a large vev,
giving rise to large masses for the mirrors.  The parameter
space of the resulting theory is large, and it is
easy to find an acceptable spectrum.

\vskip3pt
Again, I must refer the reader to ref.~\dinenelson\ for a
complete treatment of this model.  Let me close this
section by summarizing its virtues (I have already stressed
its drawbacks):

\pointbegin
DSB

\point
SU(2) $\times$  U(1) can be broken in an acceptable way; achieving
this requires additional fields, and the simplest possibility
we have listed above may be nearly the only one.

\point
The model gives adequate squark and slepton degeneracies.

\point
There are no new sources of $CP$ violation in the low energy
theory, and thus no problem with $d_n$ or $d_e$.

\point
There are no dangerous axions or goldstone bosons.

\point
All couplings are small to high energies.

\point
One can unify SU(3) $\times$ SU(2) $\times$ U(1) (at least in principle).

\point
The superpotential is the most general one allowed by
the gauge symmetries and a set of discrete symmetries.

\vskip3pt
\REF\preskill{ J. Preskill, S. P. Trivedi, F.
Wilczek, M. B. Wise,  Nucl. Phys. {\bf B363} (1991) 207.}
There is a great deal of room for further work on these
models.  One would certainly like to construct examples
which are less baroque and with a smaller degree of fine
tuning.  One might also like to find examples in which
the supersymmetry-breaking minimum is the global
minimum.  Even within this model, one would like
to further explore the parameter space, particularly
with regards to the question of SU(2) $\times$ U(1) breaking.
Finally, there are a number of cosmological issues one
would like to examine.  For example,
the gravitino mass is $10$'s
of eV.  This may pose problems for nucleosynthesis,
unless they are diluted by decays of neutralinos.
There are domain walls, though in the particular example
of ref.~\dinenelson, they disappear by the mechanism of
ref.~\preskill.  Finally, the model contains massive, long-lived
particles.

\chapter{Conclusions}

There are, as we noted earlier, four known ways to
understand the problem of squark degeneracy.  Only
one, the dilaton-driven scenario in string theory,
leads to assumptions precisely like those usually
made in the MSSM.  Two others have been explored
here:  the possibility of non-Abelian, gauged flavor
symmetries, and DSB at low energies.  Both
of these offer
alternatives to what has become the
MSSM ideology; there is still much
work to be done in exploring their phenomenology.

\bigskip
\centerline{\bf Acknowledgements}

I wish to thank my collaborators, A. Kagan, R. Leigh, and A. Nelson
for their insights into the questions discussed here.  I would
also like to thank G. Kane for several important comments.
This work was supported in part by the U.S. Department of Energy.
\endpage
\refout
\end